\documentclass{article}
\usepackage{eurosym}
\usepackage{amsfonts}
\usepackage{amsmath}
\usepackage{amsmath,amsfonts,amsthm,amssymb,amscd}
\usepackage{latexsym}
\usepackage[pdftex]{graphicx}
\usepackage{verbatim}
\usepackage{url}
\usepackage{float}
\usepackage[margin=2.25cm]{geometry}
\usepackage[titletoc]{appendix}
\usepackage{hyperref}
\usepackage{endnotes,refcount}

\input{tcilatex}
\begin{document}

\title{Automated Market-Making for Fiat Currencies}
\author{Alex Lipton and Artur Sepp \\
Sila and Sygnum Bank}
\maketitle

\begin{abstract}
We present an automated market-making (AMM) cross-settlement mechanism for
digital assets on interoperable blockchains, focusing on central bank
digital currencies (CBDCs) and stable coins. We develop an innovative
approach for generating fair exchange rates for on-chain assets consistent
with traditional off-chain markets. We illustrate the efficacy of our
approach on realized FX rates for G-10 currencies.
\end{abstract}

\section{Introduction}

Traditionally, many essays on quantitative finance start with an obligatory
reference to the celebrated Black-Scholes paper; see \cite{Black1973}.
However, as Bob Dylan put it: \textquotedblleft For the times they are
a-changin'.\textquotedblright\ Nowadays, to be \textit{au courant} with
research, one needs to start with referring to the seminal paper by Nakamoto
describing the Bitcoin protocol; see \cite{Nakamoto2008}. Publication of
this short paper by its anonymous author (or authors) started the blockchain
revolution; see \cite{Lipton2021}. However, despite its technical
brilliance, the Bitcoin protocol is limited to moving the underlying token,
called BTC, from one anonymous address to the next consistently and
robustly. Ethereum, Cardano, Polkadot, and Solana expanded Bitcoin horizons
by building a Consensus-as-a-Sevice (CaaS) machine capable of handling the
so-called smart contracts.

Over the last two years, Decentralised finance (DeFi) has seen remarkable
growth and quickly established itself as one of the first genuine
\textquotedblleft killer apps\textquotedblright\ for smart contracts; see 
\cite{Schar2021}. Using DeFi, agents can create precisely tailored and
highly complex economic arrangements and execute them automatically without
central intermediaries or other trusted parties. As a result, a DeFi-based
financial system will be more robust, inclusive, and equitable than its
ossified centralized version. Undeniably, the most valuable tools for DeFi
are automated market-making (AMM) protocols. The corresponding protocols
power all decentralized exchanges (DEXs), which gradually replace
conventional centralized exchanges relying on traditional market-making
techniques.

Of course, DeFi crypto platforms are not past teething problems. For
instance, recently, Poly Network lost approximately \$610 million worth of
crypto assets when the hacker targeted a vulnerability in the digital
interoperability contract. Eventually, the hacker returned the fund, but not
before making her point was made loud and clear.

One of the most exciting instruments using smart contracts are the so-called
Central Bank Digital Currencies (CBDCs) and Stable Coins (SCs), which map
fiat currencies on blockchains; see \cite{Liptonetal2021}. Other exciting
instruments are the so-called Non-Fungible Tokens (NFTs). Currently, CBDCs
are still being conceptualized and developed by several central banks, which
act in isolation. The ability to trade different CBDCs and SCs against each
other is critical for their adoption. While CBDCs largely remain on the
drawing board, SCs, such as Tether, USDC, Dai, and numerous others, are
well-developed and implemented as tokens on their underlying blockchains,
for example, ERC-20 tokens in Ethereum. SCs implemented on the same
blockchain can be swapped by using a smart contract. CBDCs implemented on
different blockchains can be swapped automatically only if the corresponding
blockchains are interoperable; see \cite{Hardjono2021, Lipton2022, WBG2021}.

Below we assume that SCs are implemented on the same blockchain, while CBDCs
are implemented on interoperable blockchains, and we develop an arbitrage
approach to make the on-chain exchange rate of CBDCs or SCs consistent with
the traditional off-chain forex (FX) markets. We argue that both on- and
off-chain operations are necessary to make the on-chain exchange rate
consistent with off-chain pricing. Finally, we apply our methodology for
analyzing hypothetical liquidity pools using actual FX exchange data of G-10
currencies.  We make our simulations as realistic as currently possible.

\section{ Decentralised Exchange (DEX) and Automated Marking Making}

Decentralized Exchanges with Automated market-making (AMM) have increased
over the past year as a central part of the DeFi eco-system to enable the
on-chain exchange of different tokens. Currently, Uniswap is the biggest
automated DEX with total locked value (TLV) of liquidity reserves of \$7.1B
( \url{https://defipulse.com/uniswap}, as of August 2021).

Conceptually DEX is owned by a pool operator. Liquidity providers post
tokens to the liquidity pool. Clients and traders use the pool liquidity to
exchange one CBDC for another. The exchange rate depends on the order size
based on the Constant Function Market Makers.

\textbf{The pool operator} owns the corresponding smart contract on a
blockchain and sets the core parameters of automated pool operations,
including the transaction fees. In addition, the pool operator implements
the constant function market-making (CFMM), which is implemented using a
smart contract, that provides the bid-ask quotes for client orders.

\textbf{Liquidity providers} deposit CBDCs to the pool. In return, liquidity
providers get fees generated by the pool trading activity. Liquidity
providers bear the price risk of pool assets similar to a buy-and-hold
investor. A liquidity provider may hedge CBDCs exposures in the pool using
off-chain instruments, in which case the P\&L of the provider is the stream
of fees generated by the pool trading activity.

\textbf{Traders and Clients} use the pool for either redeeming tokens from
the pool or depositing tokens to the pool in a one-sided transaction. We
assume that each client transaction of buying from the pool is charged with
the proportional pool fee.

The on-chain operations are implemented using the CFMM that assigns bid and
ask prices of two coins in the pool based on the size of the trades. As we
show, the marginal exchange rate for a trade is proportional to the
consumption of the pool liquidity. Therefore, traders and clients must
incentivize pool liquidity providers with fees accrued to the pool and
shared between the liquidity providers of the pool.

\textbf{Pool arbitrageurs} enforce the rebalancing of the pool when the CFMM
indicative bid-ask quotes are outside some thresholds. Next, we find the
optimal thresholds for a set of CFMM rules.

The pool arbitrageur implements both on-chain and off-chain transactions
intending to keep his total balance between assets and liabilities zero. The
pool arbitrageur is crucial for "price-discovery" of the on-chain pricing.
Pool operators may use designated arbitrageurs incentivized with lower fees.

\section{ Constant Function Market Makers (CFMMs)}

For clarity, we denote the EURUSD FX spot rate by $p_{t}$ and assume that $%
p_{0}=1.25$. We assume that the AMM operator creates a pool with with
notional $p_{0}N_{0}$ USDC and $N_{0}$ EUDC. The initial pool balances in
USDC, $x_{0}$, and EUDC $y_{0}$ are set respectively by: 
\begin{equation}
x_{0}=p_{0}N_{0},\ y_{0}=N_{0}.
\end{equation}

The redemption and deposits of tokens from the AMM pool is determined by the
CFMM function, also known as the pool invariant, in the following way: 
\begin{equation}
F(x_{1},y_{1};x_{0},y_{0},\Theta )=0,  \label{eq:cfmm}
\end{equation}%
where $x_{1}$ and $y_{1}$ is the amount of USDC and EUDC tokens after a
transaction, respectively; $x_{0}$ and $y_{0}$ is the amount of USDC and
EUDC tokens right before the transaction. $\Theta $ is the set of constant
set of pool parameters, including the proportional fee rate $\epsilon $.
After each transaction, pool balances $x_{0}$ and $y_{0}$ are updated by $%
x_{1}$ and $y_{1}$, respectively.

\subsection{Buying/Redeeming USDC}

Buying $\Delta x$ USDC tokens from the pool involves the redemption from the
USDC pool by deposing $(1-\epsilon )\Delta y$ tokens to the EUDC pool. The
AMM applies the CFMM (\ref{eq:cfmm}) to determine the amount of USDC $\Delta
x$ redeemed in exchange for $\Delta y$ EUDC deposited as follows: 
\begin{equation}
F^{(-,+)}(x_{0}-\Delta x,y_{0}+(1-\epsilon )\Delta y)=0.  \label{eq:cfmm_xy}
\end{equation}

Buying USDC from the pool can be done in two ways. First, the trader can
redeem known amount of $\Delta x$ USDC from pool by depositing yet unknown
amount of $\Delta \tilde{y}$ EUDC, which is found using Eq (\ref{eq:cfmm_xy}%
) with given $\Delta x$. Second, the trader can deposit a known amount of $%
\Delta y$ EUDC and redeems $\Delta \tilde{x}$ amount of USDC, which is set
using Eq (\ref{eq:cfmm_xy}) with given $\Delta y$. We summarize these
operations in Table (\ref{tb:xy}).

\begin{table}[]
\begin{tabular}{|l|l|l|l|l|}
\hline
Quote & Order & USDC balance & EUDC balance & Op for Eq(\ref{eq:cfmm_xy}) \\ 
\hline
Ask EUDC & Buy USDC/\textbf{Sell EUDC} & \textbf{redeem $\Delta x$ USDC} & 
deposit $\Delta \tilde{y}$ EUDC & $\Delta x\rightarrow\Delta \tilde{y}$ \\ 
\hline
Bid USDC & \textbf{Buy USDC}/Sell EUDC & redeem $\Delta \tilde{x}$ USDC & 
\textbf{deposit $\Delta y$ EUDC} & $\Delta y\rightarrow\Delta \tilde{x}$ \\ 
\hline
\end{tabular}%
\caption{Operations for buying USDC from the pool}
\label{tb:xy}
\end{table}

\subsection{ Buying/Redeeming EUDC}

The AMM will apply the CFMM to find the amount of EUDC $\Delta y $ redeemed
in exchange for depositing $\Delta x$ USDC using CFMM (\ref{eq:cfmm}) as
follows: 
\begin{equation}
F^{(+,-)}(x_{0}+(1-\epsilon )\Delta x,y_{0}-\Delta y)=0.  \label{eq:cfmm_yx}
\end{equation}

By analogy, the trader can redeem given amount of $\Delta y$ EUDC by
depositing the amount of $\Delta \tilde{x}$ USDC found using (\ref%
{eq:cfmm_yx}). Alternatively, the trader can deposit given amount of $\Delta
x$ USDC to redeem the amount of $\Delta \tilde{y}$ EUDC set using Eq. (\ref%
{eq:cfmm_yx}). We summarize the operations in Table (\ref{tb:yx}).

\begin{table}[]
\begin{tabular}{|l|l|l|l|l|}
\hline
Quote & Order & USDC balance & EUDC balance & Op for Eq(\ref{eq:cfmm_yx}) \\ 
\hline
Ask USDC & Buy EUDC/\textbf{Sell USDC} & deposit $\Delta \tilde{x}$ USDC & 
\textbf{redeem $\Delta y$ EUDC} & $\Delta y \rightarrow \Delta \tilde{x}$ \\ 
\hline
Bid EUDC & \textbf{Buy EUDC}/Sell USDC & \textbf{deposit $\Delta x$ USDC} & 
redeem $\Delta \tilde{y}$ EUDC & $\Delta x \rightarrow \Delta \tilde{y}$ \\ 
\hline
\end{tabular}%
\caption{Operations for buying EUDC}
\label{tb:yx}
\end{table}

\subsection{ Bid/Ask Matching}

In Table (\ref{tb:ba}), we aggregate the operations in Tables (\ref{tb:xy})
and (\ref{tb:yx}) to present the conventional matching of order book.

\begin{table}[]
\begin{tabular}{|l|l|l|}
\hline
Quote & USDC & EUDC \\ \hline
Bid & \textbf{Buy USDC}/Sell EUDC: $F^{(-,+)}:\Delta y \rightarrow \Delta 
\bar{x}$ & \textbf{Buy EUDC}/Sell USDC: $F^{(+,-)}:\Delta x \rightarrow
\Delta \bar{y}$ \\ \hline
Ask & Buy EUDC/\textbf{Sell USDC}: $F^{(+,-)}:\Delta y \rightarrow \Delta 
\bar{x}$ & Buy USDC/\textbf{Sell EUDC}: $F^{(-,+)}:\Delta x \rightarrow
\Delta \bar{y}$ \\ \hline
\end{tabular}%
\caption{Bid/Ask quoting as function of order sizes using CFMM}
\label{tb:ba}
\end{table}

\section{ Representative examples}

We now present the most important types of CFMMs; among numerous others, see 
\cite{Angeris2019, Egorov2019}.

\subsection{ Constant sum function}

Constant sum rule specifies the following invariant for CFMM (\ref{eq:cfmm}%
): 
\begin{equation}
F(x_{1},y_{1})\equiv x_{1}+sy_{1}-\Sigma _{0}=0.  \label{eq:cs1}
\end{equation}%
Here and below $s$ is the fixed level of \textquotedblleft
equilibrium\textquotedblright\ conversion rate of $y$ tokens to $x$ tokens,
and $\Sigma _{0}=x_{0}+sy_{0}$.

For redemption of USDC, Eq(\ref{eq:cfmm_xy}) simplifies to: 
\begin{equation}
\Delta x=s(1-\epsilon )\Delta y,  \label{eq:cs3}
\end{equation}%
and for for redemption of EUDC, Eq(\ref{eq:cfmm_xy}) becomes: 
\begin{equation}
(1-\epsilon )\Delta x=s\Delta y.  \label{eq:cs5}
\end{equation}

In Table (\ref{tb:bacs}), we summarise bid/ask matching as defined in Table (%
\ref{tb:ba}). 
\begin{table}[]
\label{tb:bacs} 
\begin{tabular}{|l|l|l|}
\hline
Quote & USDC & EUDC \\ \hline
Bid & Redeem USDC: $F^{(-,+)}/\Delta y = (1-\epsilon)s $ & Redeem EUDC: $%
F^{(+,-)}/\Delta x = \frac{(1-\epsilon)}{s}$ \\ \hline
Ask & Deposit USDC: $F^{(+,-)}/\Delta y = \frac{s}{(1-\epsilon)} $ & Deposit
EUDC: $F^{(-,+)}/\Delta x = \frac{1}{s(1-\epsilon)}$ \\ \hline
\end{tabular}%
\caption{Bid/Ask order quoting for constant sum function.}
\end{table}

\begin{figure}[]
\begin{center}
\includegraphics[width=1.\textwidth, angle=0]
{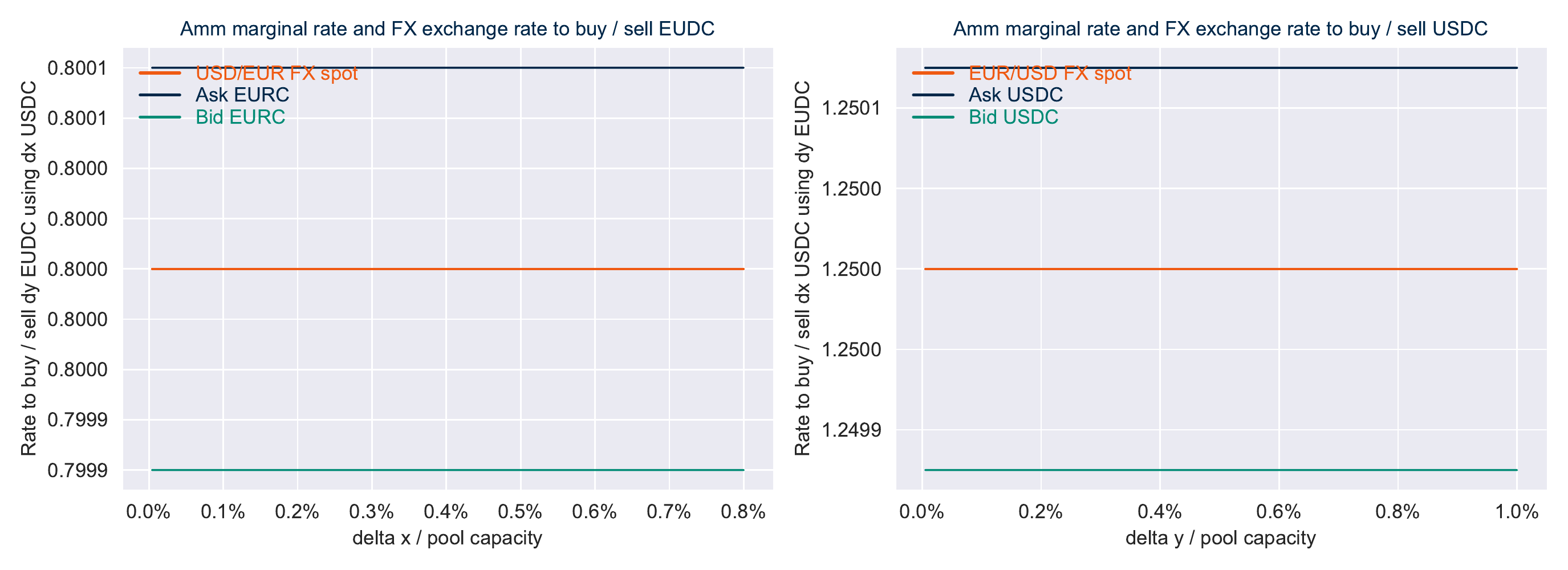}%
\vspace*{-\baselineskip}
\end{center}
\par
\vspace*{-\baselineskip}
\caption{Left: the AMM rate to buy/sell EUDC as function of $\Delta x$ USDC, 
$\Delta x=[1,2,..,100]$, with x-axis is pool utilization rate $\Delta x /
x_{0}$. Right: the AMM rate to buy/sell USDC as function of $\Delta y$ EUDC, 
$\Delta y=[1,2,..,100]$, with y-axis is pool utilization rate $\Delta y /
y_{0}$ Initial parameters $s=p_{0}=1.25$ (EUR/USD FX spot) and $1/p_{0}=0.80$
(USD/EUR FX spot), $N_{0}=10000$, $\protect\epsilon=1bp$.}
\label{fig:csum}
\end{figure}

In Figure (\ref{fig:csum}), we show the bid ask rates for USDC and EUDC
implied by the AMM. It is clear that the constant sum rule does not depend
on the order size, so that it is less realistic for AMM.

\subsection{ Constant product function}

Constant product function specifies the following invariant for CFMM (\ref%
{eq:cfmm}): 
\begin{equation}
F(x_{1},y_{1})\equiv sx_{1}y_{1}-\Pi _{0}=0.  \label{eq:cp1}
\end{equation}%
The AMM will apply the CFMM to find the amount of USDC $\Delta x$ redeemed
by the trader in exchange for depositing $\Delta y$ EUDC as follows: 
\begin{equation}
F^{(-,+)}\equiv \left( x_{0}-\Delta x\right) \left( y_{0}+(1-\epsilon
)\Delta y\right) =\Pi _{0}/s.
\end{equation}%
Accordingly, for the operation of redeeming a given amount of USDC $\Delta x$
the amount of EUDC received $\Delta \bar{y}$ is found by: 
\begin{equation}
\Delta \bar{y}=\frac{1}{(1-\epsilon )}\left( \frac{\Pi _{0}}{s\left(
x_{0}-\Delta x\right) }-y_{0}\right) .  \label{eq:x_for_y_with_dx}
\end{equation}%
For the operation of depositing a given amount of EUDC $\Delta y$, the
amount of USDC $\Delta \bar{x}$ is set by: 
\begin{equation}
\Delta \bar{x}=x_{0}-\frac{\Pi _{0}}{s\left( y_{0}+(1-\epsilon )\Delta
y\right) }.  \label{eq:x_for_y_with_dy}
\end{equation}

Similarly, we apply the invariant for redeeming a given amount of EUDC $%
\Delta y$. In Table (\ref{tb:fcprod}), we provide the summary for order
matching.

\begin{table}[]
\begin{tabular}{|l|l|l|}
\hline
Quote & USDC & EUDC \\ \hline
Bid & Redeem: $F^{(-,+)}/\Delta y = \frac{x_{0}}{\Delta y} - \frac{\Pi_{0} }{%
s\Delta y\left(y_{0} + (1-\epsilon)\Delta y\right)}$ & Redeem: $%
F^{(+,-)}/\Delta x = \frac{1}{(1-\epsilon)\Delta x}\left(\frac{\Pi_{0}}{%
s\left(x_{0} - \Delta x \right) }-y_{0}\right) $ \\ \hline
Ask & Deposit: $F^{(+,-)}/\Delta y = \frac{1}{(1-\epsilon)\Delta y} \left( 
\frac{\Pi_{0}}{s\left(y_{0} - \Delta y\right) }-x_{0}\right)$ & Deposit: $%
F^{(-,+)}/\Delta x = \frac{y_{0}}{\Delta x} - \frac{\Pi_{0} }{s\Delta
x\left(x_{0} + (1-\epsilon)\Delta x\right)}$ \\ \hline
\end{tabular}%
\caption{Bid/Ask order matching for constant product function}
\label{tb:fcprod}
\end{table}

\begin{figure}[]
\begin{center}
\includegraphics[width=1.\textwidth, angle=0]
{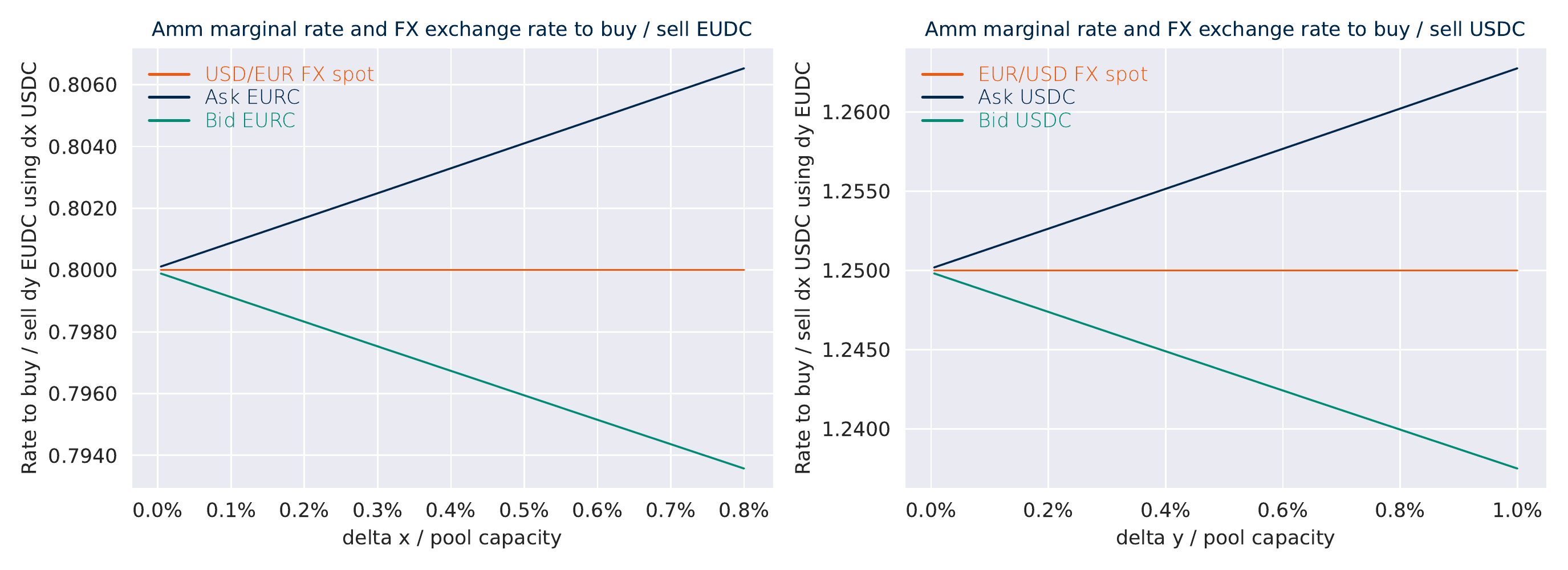}%
\vspace*{-\baselineskip}
\end{center}
\par
\vspace*{-\baselineskip}
\caption{Left: the AMM rate to buy/sell EUDC implied by the constant product
AMM as function of pool utilization rate $\Delta x / x_{0}$. Right: the AMM
rate to buy/sell USDC as function of pool utilization $\Delta y / x_{0}$.
Initial parameters: $p_{0}=1.25$, $s=p_{0}$, $\Pi_{0}=10000$, $\protect%
\epsilon=1bp$.}
\label{fig:fcprod}
\end{figure}

In Figure (\ref{fig:fcprod}), we show the bid-ask rates for USDC and EUDC
implied by the constant product AMM. The constant product rule makes large
orders prohibitively expensive.

\subsection{ Mixed rule function}

Mixed rule function specifies the following invariant for CFMM (\ref{eq:cfmm}%
): 
\begin{equation}
F(x_{1},y_{1})\equiv \left( \frac{\Pi _{0}}{sx_{1}y_{1}}-1\right) -\alpha
\left( \frac{x_{1}+sy_{1}}{\Sigma _{0}}-1\right) =0,  \label{eq:mr1}
\end{equation}%
where the pool composition is defined by sum and product: 
\begin{equation}
x_{0}+sy_{0}=\Sigma _{0}\ ,\ sx_{0}y_{0}=\Pi _{0}.
\end{equation}%
It is clear that the mixed rule is symmetric under the change $%
(x,sy)\rightarrow (sx,y)$.

If $x_{1}$ is given, Eq. (\ref{eq:mr1}) is solved for $\tilde{y}%
_{1}^{(\alpha )}(x_{1})$: 
\begin{equation}
y_{1}^{(\alpha )}(x)=\frac{1}{2\tilde{\alpha}s}\left( -\left( (1-\alpha )+%
\tilde{\alpha}x\right) +\sqrt{D}\right) ,\ D=\left( (1-\alpha )+\tilde{\alpha%
}x\right) ^{2}+\frac{4\tilde{\alpha}\Pi }{x},  \label{eq:mr2_xy}
\end{equation}%
where $\tilde{\alpha}\equiv \alpha /\Sigma $. If $y_{1}$ is given, Eq. (\ref%
{eq:mr1}) is solved for $\tilde{x}_{1}^{(\alpha )}(y_{1})$:

\begin{equation}
x_{1}^{(\alpha )}(y)=\frac{1}{2\tilde{\alpha}}\left( -\left( (1-\alpha )+%
\tilde{\alpha}sy\right) +\sqrt{D}\right) ,\ D=\left( (1-\alpha )+\tilde{%
\alpha}sy\right) ^{2}+\frac{4\tilde{\alpha}\Pi }{sy}.  \label{eq:mr2_yx}
\end{equation}

To redeem given $\Delta x$ USDC from the pool by depositing $\Delta \tilde{y}
$ EUDC, we solve Eq (\ref{eq:mr2_xy}) for $\tilde{y}_{1}$ using $%
x_{1}=x_{0}-\Delta x$ and set: 
\begin{equation}
\Delta x=x_{0}-x_{1}\ ,\Delta \tilde{y}=\frac{\left( \tilde{y}%
_{1}-y_{0}\right) }{(1-\epsilon )}.  \label{eq:mr_us1}
\end{equation}

By analogy we consider the other 3 transactions and summarise the outputs in
Table (\ref{tb:fcprod1}). 
\begin{table}[]
\begin{tabular}{|l|l|l|}
\hline
Quote & USDC & EUDC \\ \hline
Bid & Redeem: $\Delta x = x_{0}- x_{1} \ , \Delta \tilde{y} = \frac{\left(%
\tilde{y}_{1} - y_{0}\right)}{(1-\epsilon)} $ & Redeem: $\Delta x = \frac{%
\left(x_{1} - x_{0}\right)}{(1-\epsilon)} , \ \Delta \tilde{y} = y_{0}- 
\tilde{y}_{1}$ \\ \hline
Ask & Deposit: $\Delta \tilde{x} = \frac{\left(\tilde{x}_{1} - x_{0}\right)}{%
(1-\epsilon)} , \ \Delta y = y_{0}- y_{1} $ & Deposit: $\Delta\tilde{x} =
x_{0} - \tilde{x}_{1} \ , \Delta y = \frac{\left(y_{1} - y_{0}\right)}{%
(1-\epsilon)} $ \\ \hline
\end{tabular}%
\caption{Bid/Ask order quoting for mixed rule AMM}
\label{tb:fcprod1}
\end{table}

\begin{figure}[]
\begin{center}
\includegraphics[width=1.\textwidth, angle=0]
{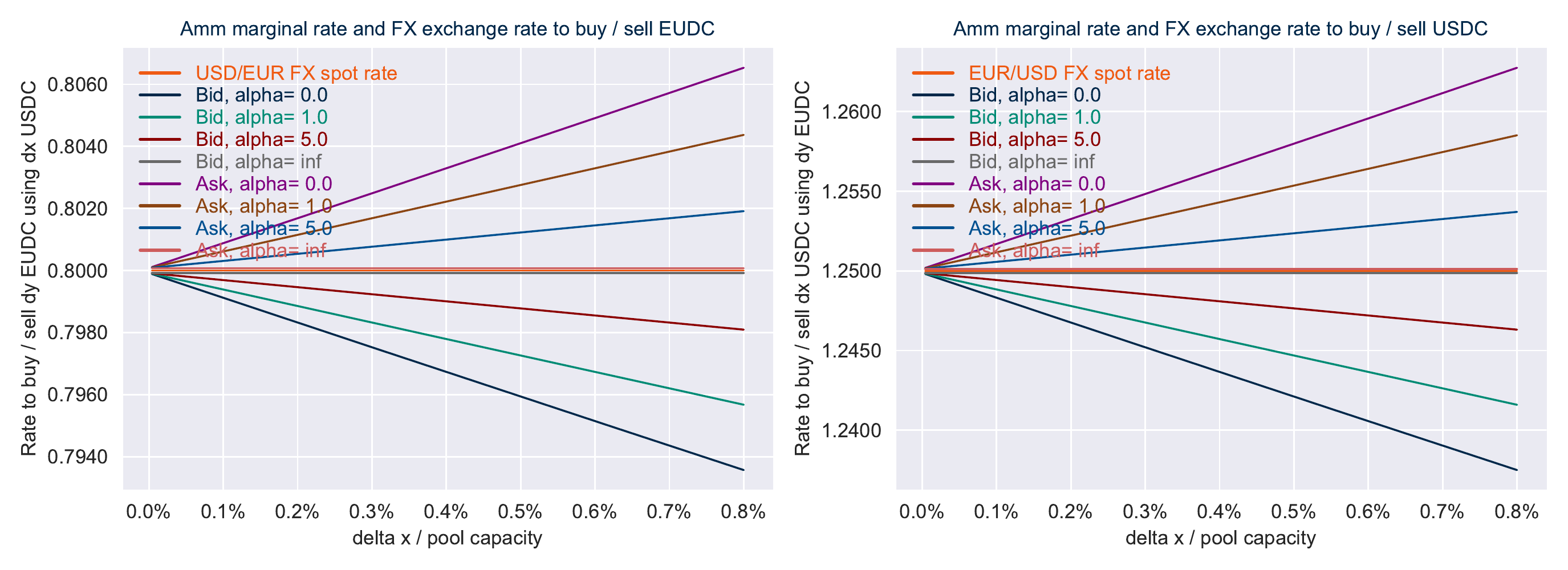}%
\vspace*{-\baselineskip}
\end{center}
\par
\vspace*{-\baselineskip}
\caption{Left: the AMM rate to buy/sell EUDC implied by the mixed rule AMM
as function of pool utilization rate $\Delta x / x_{0}$. Right: the AMM rate
to buy/sell USDC as function of pool utilization $\Delta y / x_{0}$. Initial
parameters: $p_{0}=1.25$, $s=p_{0}$, $\Pi_{0}=10000$, $\protect\epsilon=1bp$%
. }
\label{fig:mixed_rule}
\end{figure}

In Figure (\ref{fig:mixed_rule}), we show the bid/ask rates as functions of
order size relative to the pool liquidity using different values of the
parameter $\alpha $. Here $\alpha =0$ and $\alpha =\infty $ correspond to
the sum and product rules, respectively. Using $\alpha $, we can control the
marginal exchange rate as a function of order size.

\section{AMM arbitrage}

The arbitrage of the pool exchange rate includes both off-chain transactions
of buying/selling FX spots and on-chain redemption/deposit of DC.

\subsection{Operations}

We start with USDC/EUDC arbitrage which can be executed if the marginal rate
of USDC/EUDC is above the FX rate. The arbitrageur will borrow USD cash to
buy EUR cash off-chain, convert it to EUDC, exchange EUDC to USDC, and
finally convert USDC back to USD cash to cover the initial USD margin to buy
EUR cash. The operations are summarised in table (\ref{tb:useu}).

\begin{table}[]
\begin{tabular}{|l|l|l|l|l|l|l|}
\hline
& Operation & Type & USD cash & EUR cash & USDC & EUDC \\ \hline
1 & Borrow USD / Buy EUR & OFF & $-P_{E/U}\Delta y$ & $\Delta y$ & 0 & 0 \\ 
\hline
2 & Convert EUR to EUDC & OFF$\rightarrow$ON & $-P_{E/U}\Delta y$ & 0 & 0 & $%
\Delta y$ \\ \hline
3 & Redeem USDC/deposit EUDC & SC & $-P_{E/U}\Delta y$ & 0 & $\Delta x$ & 0
\\ \hline
4 & Convert USDC to USD & ON$\rightarrow$OFF & $\Delta x -P_{E/U}\Delta y$ & 
0 & 0 & 0 \\ \hline
\end{tabular}%
\caption{Operations for arbitrage high USDC/EUDC rate implied by the pool. $%
P_{E/U}$ is the spot FX EUR/USD rate. Type OFF is off-chain, ON is on-chain,
and SC is smart contract}
\label{tb:useu}
\end{table}

Conversely, EUDC/USDC can be executed if the marginal rate of EUDC/USDC is
above the FX rate. The arbitrageur will borrow EUR cash to buy USD cash
off-chain, convert it to USDC, exchange USDC to EUDC, and finally convert
EUDC back to EUR cash to cover the initial EUR margin to buy USD cash. We
can summarise the operations similarly to Table (\ref{tb:useu}).

Pool arbitrageurs are necessary to adjust the pool compositions so that the
AMM rate is fair and attractive for clients. AMM operators may incentivize
arbitrageurs to enforce the pool's price discovery through a sequence of on-
and off-chain transactions.

\subsection{ Formulation}

\subsubsection{ USDC arbitrage}

For USDC arbitrage, we use the CFMM (\ref{eq:cfmm_xy}) to deposit given
amount $\Delta y$ EUDC redeem $\Delta x$ USDC and deposit EUDC. We use
operations in Table (\ref{tb:useu}) to derive the arbitrage P\&L as function
of $\Delta y$: 
\begin{equation}
\begin{split}
& \Omega (\Delta y;p_{1},x_{0},y_{0})=\Delta \bar{x}-p_{1}\Delta y, \\
& \text{s.t.}\ F(x_{0}-\Delta \bar{x},y_{0}+(1-\epsilon ^{\prime })\Delta
y)=0,
\end{split}
\label{eq:useu1}
\end{equation}%
where $p_{1}$ is EUR/USD spot rate.

Here we assume that the $\epsilon ^{\prime }$ is the fee paid by
arbitrageur. The optimal arbitrage trade size is: 
\begin{equation}
\Delta y^{\ast }=\text{argmax}_{\Delta y}\Omega (\Delta y;p_{1},x_{0},y_{0}),
\label{eq:useu2}
\end{equation}%
and the rebalancing condition is $\Delta y^{\ast }>0$.

\subsubsection{ EUDC arbitrage}

For EUDC arbitrage, we use the CFMM (\ref{eq:cfmm_yx}) to deposit given
amount $\Delta x$ USDC to redeem $\Delta \bar{y}$ USDC. We derive the
arbitrage P\&L as function of $\Delta x$ as follows: 
\begin{equation}
\begin{split}
& \Omega (\Delta x;p_{1},x_{0},y_{0})=p_{1}\Delta \bar{y}-\Delta x, \\
& \text{s.t.}\ F(x_{0}+(1-\epsilon ^{\prime })\Delta x,y_{0}-\Delta \bar{y}%
)=0,
\end{split}
\label{eq:euus1}
\end{equation}%
where $p_{1}$ is EUR/USD spot rate. Here $\epsilon ^{\prime }$ is the fees
to add EUCD to the AMM pool with the total fees incurred by the arbitrage
for all 4 operations in Table (\ref{tb:useu}).

The optimal arbitrage trade size is: 
\begin{equation}
\Delta x^{\ast }=\text{argmax}_{\Delta x}\Omega (\Delta x;p_{1},x_{0},y_{0}).
\label{eq:euus2}
\end{equation}

\subsection{ Arbitrage with Constant product rule}

For USDC arbitrage, we use Eq(\ref{eq:useu1}) with Eq (\ref%
{eq:x_for_y_with_dy}) for $\Delta \bar{x}$: 
\begin{equation}
\Omega (\Delta y)=\left( x_{0}-\frac{\Pi _{0}}{s\left( y_{0}+(1-\epsilon
^{\prime })\Delta y\right) }\right) -p_{1}\Delta y,
\end{equation}%
so that 
\begin{equation}
\Delta y^{\ast }=\left( \sqrt{\frac{\Pi _{0}}{s(1-\epsilon ^{\prime })p_{1}}}%
-\frac{y_{0}}{(1-\epsilon ^{\prime })}\right) .  \label{eq:cpdy}
\end{equation}

Similarly for EUDC Arbitrage, we obtain the optimal trade size:

\begin{equation}
\Delta x^{\ast }=\left( \sqrt{\frac{p_{1}\Pi _{0}}{s(1-\epsilon ^{\prime })}}%
-\frac{x_{0}}{(1-\epsilon ^{\prime })}\right) .  \label{eq:cpdx}
\end{equation}

Rebalancing condition $\Delta y^{\ast }>0$ and $\Delta x^{\ast }>0$ imply
the following bands: 
\begin{equation}
\frac{1}{(1-\epsilon ^{\prime })}\frac{x_{0}}{y_{0}}<p_{1}<(1-\epsilon
^{\prime })\frac{x_{0}}{y_{0}}.
\end{equation}

\subsection{ Arbitrage with Mixed Rule}

For USDC arbitrage, we use Eq(\ref{eq:useu1}) with $\Delta \bar{x}$: 
\begin{equation}
\Omega (\Delta y)=\left( x_{0}-x_{1}^{(\alpha )}(y_{0}+(1-\epsilon ^{\prime
})\Delta y)\right) -p_{1}\Delta y,
\end{equation}%
where $x_{1}^{(\alpha )}(y)$ solves Eq (\ref{eq:mr2_yx}).

The optimality condition is 
\begin{equation}
\Omega ^{\prime }(\Delta y)=-(1-\epsilon ^{\prime })\partial
_{y}x_{1}^{(\alpha )}(y)-p_{1},
\end{equation}%
where $y=y_{0}+(1-\epsilon ^{\prime })\Delta y$.

Accordingly, we solve for $y^{\ast }$ using Newton-Raphson: 
\begin{equation}
y_{n+1}^{\ast }=y_{n}^{\ast }-\frac{\partial _{y}x_{1}^{(\alpha
)}(y_{n}^{\ast })+p_{1}/(1-\epsilon ^{\prime })}{\partial
_{yy}x_{1}^{(\alpha )}(y_{n}^{\ast })},
\end{equation}%
and take 
\begin{equation}
\Delta y^{\ast }=\frac{\left( y_{n}^{\ast }-y_{0}\right) }{1-\epsilon
^{\prime }}.
\end{equation}

Similarly, we derive the optimal trade size for EUDC arbitrage.

\begin{figure}[]
\begin{center}
\includegraphics[width=1.\textwidth, angle=0]
{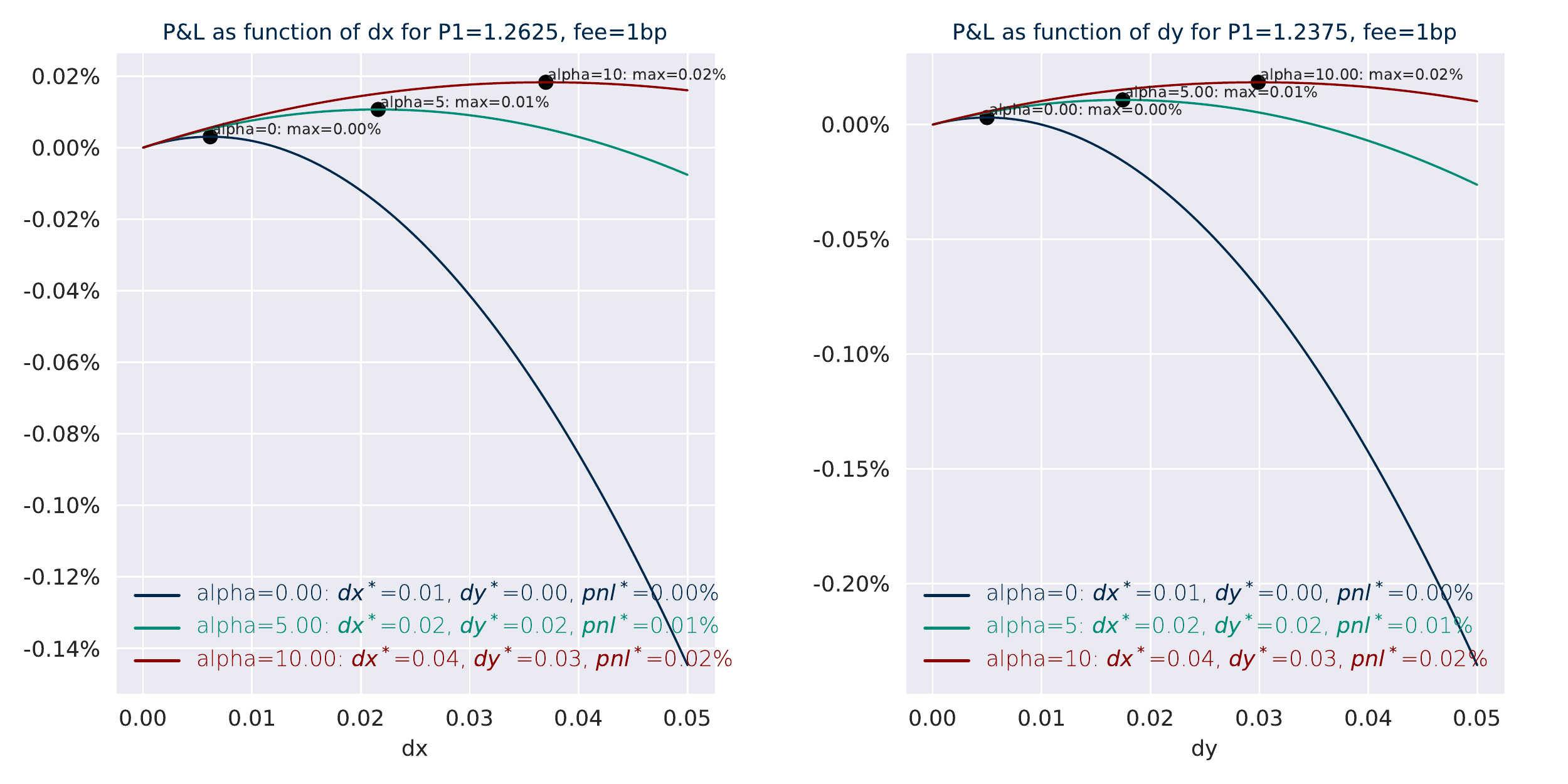}%
\vspace*{-\baselineskip}
\end{center}
\par
\vspace*{-\baselineskip}
\caption{Optimal P\&L and rebalancing as function of $\protect\alpha$ using
fee $1$bp}
\label{fig:opt_pnl_alpha}
\end{figure}

In Figure (\ref{fig:opt_pnl_alpha}), we show the optimal arbitrage P\&L and
rebalancing as function of $\alpha$. The constant product rule is the most
expensive for arbitrageurs, so the optimal trade must be small. High alpha
values produce smaller marginal costs of using AMM so that the arbitrageurs
can trade in larger sizes with higher profit potential. 

\section{ Pool simulations using FX Data}

We apply the actual FX data, assuming it is representative of DCs, for
simulation of the dynamics of the pool, pool spreads, and P\&L of liquidity
providers.

\subsection{ FX Data and Normalization}

We use FX data downloaded from the Dukascopy Bank SA platform. The FX data
represent one-minute open-high-low-close quotes and traded volumes for both
bid and ask trades. These bid/ask data are well suited for our purposes
because we need to apply both the price and the volume data for realistic
simulations of the FX pools. We chose one-minute bars because they would be
representative of an average block mining time.

We consider each 24-hour trading session a complete life-cycle of the FX
pool, independent of previous sessions. At the start of each trading
session, the initial pool balances are reset, and, at the end of the trading
session, selected variables of the pool activity are recorded. We use the
last three years of the FX data as the most representative. As a result, we
obtain $780$ ($=3\times260$) independent realizations of daily pool
variables for each FX pool. Each 24-hour session has a total of $1440$ ($%
=24\times60$) data points.

We use 10 FX pairs for G-10 currencies and Chinese yuan including the
following FX pairs: 'EURUSD', 'GBPUSD', 'USDJPY', 'USDCHF', 'AUDUSD',
'NZDUSD', 'USDCAD', 'USDNOK', 'USDSEK', 'USDCNH'.

For each trading session, we normalize the 1-minute close prices as follows: 
\begin{equation}
P^{mid}(t)=\frac{1}{2}\left( P^{bid}(t)+P^{ask}(t)\right) ,\ \ \ p^{i}\left(
t\right) =\frac{P^{i}\left( t\right) }{P^{mid}\left( 0\right) },\ \ \
i=mid,bid,ask,  \label{eq:d1}
\end{equation}%
where capital $P$ stands for natural prices and small $p$ stands for
normalized prices.

The natural bid and ask volumes represent the price-weighted volumes of sell
and buy trades, respectively. We normalize the volume data for each session
to add up to one as follows:

\begin{equation}
\begin{array}{c}
\hat{V}=\left( \sum_{t}V^{bid}(t)+\sum_{t}V^{ask}(t)\right) ,\ v^{bid}(t)=%
\frac{1}{\hat{V}}V^{bid}(t),\ v^{ask}(t)=\frac{1}{\hat{V}}V^{ask}(t)\ .%
\end{array}
\label{eq:d2}
\end{equation}%
where capital $V$ stands for natural volumes and small $v$ stands for
normalized volumes. It follows that normalized bid and ask volumes represent
the unit-based buy and sell trades. Thus, the total amount of daily volume
sums up to $100\%$.

\subsection{ Pool Specification}

For each FX pair, we set $y$ to represent the pool balances for the foreign
DC and $x$ represent the pool balances for the domestic DC. The price $%
p^{mid}(t)$ is set to be the market rate of exchange of one unit of the
foreign currency to $p^{mid}(t)$ units of the domestic currency. For
example, for the 'EURUSD' FX pair, EUR (EUDC) is the foreign currency with
the pool balances equal $y$, and USD (USDC) is the domestic currency with
the pool balance equal $x$.

The trade volumes to redeem (buy) the foreign DC from the pool balances $y$
are set to $\Delta y^{bid}(t) = v^{bid}(t)$ with the matched volumes to
deposit (sell) the domestic DC to pool balances $x$ are set to $\Delta 
\overline{x}^{ask}(t)$, which is the output from the CFMM. Conversely, the
trade volumes to redeem (buy) the domestic DC from the pool balances $x$ is
set to $\Delta x^{bid}(t) = v^{ask}(t)$, with the matched volumes to deposit
(sell) the foreign DC to the pool are set to $\Delta \overline{y}^{ask}(t)$,
which is the output from the CFMM.

At each trading session, the pool balances are initialized as follows: 
\begin{equation}
x_{0}=1,\ y_{0}=1,\ s=1,\ \Pi _{0}=sx_{0}y_{0}=1,\ \Sigma
_{0}=x_{0}+sy_{0}=2.  \label{eq:d3}
\end{equation}

Given that the initial FX rate for $y$ prices in $x$ is normalized to unity,
the starting value of the pool (also known as total value locked or TLV) is $%
1$. Because the normalized volumes add up to $1$, we make an explicit
assumption that the daily turnover of the pool is $100\%$ (without
accounting for intraday FX rate fluctuations) for each of the DC pools.

The actual pool capacity and pool turnover are expected to be varying on
daily basis, so we keep the pool turnover as an endogenous variable set to
unity. We assume that single liquidity provides the pool liquidity provider
so that the pool accrued fees and the P\&L correspond to the liquidity
provider with $100\%$ ownership of the pool.

\subsection{ Intraday Simulation of the Pool}

For each trading session, the time stamps $t_{n}$ represent one-minute
intervals. The 1-minute realizations of the intraday dynamics of the pool
are simulated as follows.

\begin{enumerate}
\item \addtocounter{enumi}{-1}

\item \textbf{Initialization.} The pool is initialized at the beginning of
each trading session using Eq (\ref{eq:d3}).

\item \textbf{Order arrivals.} We assume that all bid and ask orders are
aggregated into the respective single bid and ask order using normalized bid
and ask trade volumes at $n$-th period. We assume one lag delay, so bid and
ask orders are processed first using the pool data at $(n-1)$-th period.

The bid volumes for foreign DC $\Delta y_{n}^{bid}=v_{n}^{bid}$ correspond
to redeeming/buying foreign DC from balances $y$ with matched deposit size $%
\Delta \bar{x}_{n}$ to domestic DC balance $x$ processed using CFMM
specified in Eq. (\ref{eq:cfmm_xy}): 
\begin{equation}
F^{(+,-)}(x_{n-1}+(1-\epsilon )\Delta \bar{x}_{n}^{ask},y_{n-1}-\Delta
y_{n}^{bid})=0,  \label{eq:ps1a}
\end{equation}%
where $\Delta y_{n}^{bid}=v_{n}^{bid}$.

The bid volumes for domestic DC $\Delta x_{n}^{bid}=v_{n}^{ask}$ correspond
to buying/redeeming domestic DC for the pool balances $x$ with matched
deposit size $\Delta \bar{y}_{n}^{ask}$ to foreign DC balance $y$ processed
using CFMM specified in Eq. (\ref{eq:cfmm_yx}): 
\begin{equation}
F^{(-,+)}(x_{n-1}-\Delta x_{n}^{bid},y_{n-1}+(1-\epsilon )\Delta \bar{y}%
_{n}^{ask})=0,  \label{eq:ps1b}
\end{equation}%
where $\Delta x_{n}^{bid}=v_{n}^{ask}$. The pool composition at the end of
period $n$ is then updated as follows: 
\begin{equation}
\begin{split}
& x_{n}=x_{n-1}-\Delta x_{n}^{bid}+\Delta \bar{x}_{n}^{ask}, \\
& y_{n}=y_{n-1}-\Delta y_{n}^{bid}+\Delta \bar{y}_{n}^{ask}.
\end{split}
\label{eq:ps2}
\end{equation}%
The marginal rate for the AMM bid/ask order fills is given by: 
\begin{equation}
p_{n}^{AMM,bid}=\frac{\Delta x_{n}^{bid}}{\Delta \bar{y}_{n}^{ask}},\
p_{n}^{AMM,ask}=\frac{\Delta \bar{x}_{n}^{ask}}{\Delta y_{n}^{bid}},
\label{eq:ps2a}
\end{equation}%
We compute the bid/ask spread as follows: 
\begin{equation}
s_{n}^{AMM,bid}=p_{n}^{mid}-p_{n}^{AMM,bid},\
s_{n}^{AMM,ask}=p_{n}^{AMM,ask}-p_{n}^{mid}.  \label{eq:ps2b}
\end{equation}

\item \textbf{Arbitrage Operation}

The arbitrage is implemented by checking the optimal arbitrage trade sizes
in Eq (\ref{eq:useu2}) and Eq (\ref{eq:euus2}) for the optimal arbitrage
trade sizes: 
\begin{equation}
\begin{split}
& \Delta y_{n}^{\ast }=\text{argmax}_{\Delta y}\Omega (\Delta
y;p_{n-1}^{mid},x_{n},y_{n}). \\
& \Delta x_{n}^{\ast }=\text{argmax}_{\Delta x}\Omega (\Delta
x;p_{n-1}^{mid},x_{n},y_{n}),
\end{split}
\label{eq:ps3}
\end{equation}%
using updated pool compositions in Eq (\ref{eq:ps2}). It is clear that
either $\Delta y_{n}^{\ast }>0$ or $\Delta x_{n}^{\ast }>0$, but not both,
so that the arbitrageur must only make a one-sided rebalancing and the
matching sizes $\Delta \bar{x}_{n}^{\ast }$ and $\Delta \bar{y}_{n}^{\ast }$
are computed using the CFMM, respectively.

Further we assume that the arbitrage immediately implements the arbitrage
adjustments (\ref{eq:ps3}) with the fill price for the FX spot transaction
given by $p_{n}^{mid}$. Thus, the arbitrage profit as specified by Eq. (\ref%
{eq:useu1}) and (\ref{eq:euus1}) for the $n$-th period is computed by: 
\begin{equation}
\begin{split}
& \Omega (\Delta y_{n}^{\ast };p_{n}^{mid},x_{n},y_{n})=\Delta \bar{x}%
_{n}^{\ast }-p_{n}^{mid}\Delta y_{n}^{\ast }, \\
& \Omega (\Delta x_{n}^{\ast };p_{n^{\prime
}}^{mid},x_{n},y_{n})=p_{n}^{mid}\Delta \bar{y}_{n}^{\ast }-\Delta
x_{n}^{\ast }.
\end{split}
\label{eq:ps4}
\end{equation}

\item \textbf{Update of the pool composition}

At the end of each $n$-th period, the pool composition is updated by
accounting for volume-driven rebalancing in Eq. (\ref{eq:ps2}) and the
arbitrage-driven rebalancing in Eq (\ref{eq:ps3}): 
\begin{equation}
\begin{split}
& x_{n}=x_{n-1}+\left( -\Delta x_{n}^{bid}+\Delta \bar{x}_{n}^{ask}\right)
+\left( \Delta x_{n}^{\ast }-\Delta \bar{x}_{n}^{\ast }\right) , \\
& y_{n}=y_{n-1}+\left( -\Delta y_{n}^{bid}+\Delta \bar{y}_{n}^{ask}\right)
+\left( \Delta y_{n}^{\ast }-\Delta \bar{y}_{n}^{\ast }\right) .
\end{split}
\label{eq:ps5}
\end{equation}

\item \textbf{Record of pool variables at the end of the trading session}

At the end of each trading session with $n=N$, we compute the following key
variables.

\begin{itemize}
\item Volume-weighted average AMM bid-ask spread in basis points is computed
by:

\begin{equation}
s_{N}=\frac{10000}{2}\left( \sum_{n^{\prime }=1}^{N}v_{n^{\prime
}}^{bid}s_{n^{\prime }}^{AMM,bid}+\sum_{n^{\prime }=1}^{N}v_{n^{\prime
}}^{ask}s_{n^{\prime }}^{AMM,ask}\right) ,  \label{eq:ps8}
\end{equation}%
where $s_{n}^{AMM,bid}$ and $s_{n}^{AMM,ask}$ are defined in Eq. (\ref%
{eq:ps2b}) and $10000$ is the basis point scaler.

\item Arbitrage profits are computed using Eq. (\ref{eq:ps4}): 
\begin{equation}
p\&l_{N}^{arb}=\sum_{n^{\prime }=1}^{N}\Omega (\Delta y_{n^{\prime }}^{\ast
};p_{n^{\prime }}^{mid},x_{n^{\prime }},y_{n^{\prime }})+\sum_{n^{\prime
}=1}^{N}\Omega (\Delta x_{n^{\prime }}^{\ast };p_{n^{\prime
}}^{mid},x_{n^{\prime }},y_{n^{\prime }}).  \label{eq:ps9}
\end{equation}

\item Total pool fees are computed using pool trades set in Eq. (\ref%
{eq:ps1a}) and (\ref{eq:ps1b}): 
\begin{equation}
f_{N}=\epsilon \left( \sum_{n^{\prime }=1}^{N}\Delta \bar{x}_{n^{\prime
}}^{ask}+\sum_{n^{\prime }=1}^{N}p_{n^{\prime }}^{mid}\Delta \bar{y}%
_{n^{\prime }}^{ask}\right) .  \label{eq:ps10}
\end{equation}

\item Total P\&L for the pool liquidity provider (with $100\%$ pool
ownership) including the impermanent loss p\&L and total fees is computed
by: 
\begin{equation}
p\&l_{N}^{lp}=\left( x_{N}+p_{N}^{mid}y_{N}\right) -\left(
x_{0}+p_{0}^{mid}y_{0}\right) +f_{N}.  \label{eq:ps11}
\end{equation}

\item Total P\&L for the pool liquidity provider with hedging including the
impermanent loss p\&L with hedging and total fees is computed by: 
\begin{equation}
p\&l_{N}^{lp-hedged}=\left( x_{N}+p_{N}^{mid}y_{N}\right) -\left(
x_{0}+p_{N}^{mid}y_{0}\right) +f_{N}.  \label{eq:ps12}
\end{equation}
\end{itemize}
\end{enumerate}

\subsection{ Illustration of Intraday dynamics}

\begin{figure}[]
\begin{center}
\includegraphics[width=1.\textwidth, angle=0]
{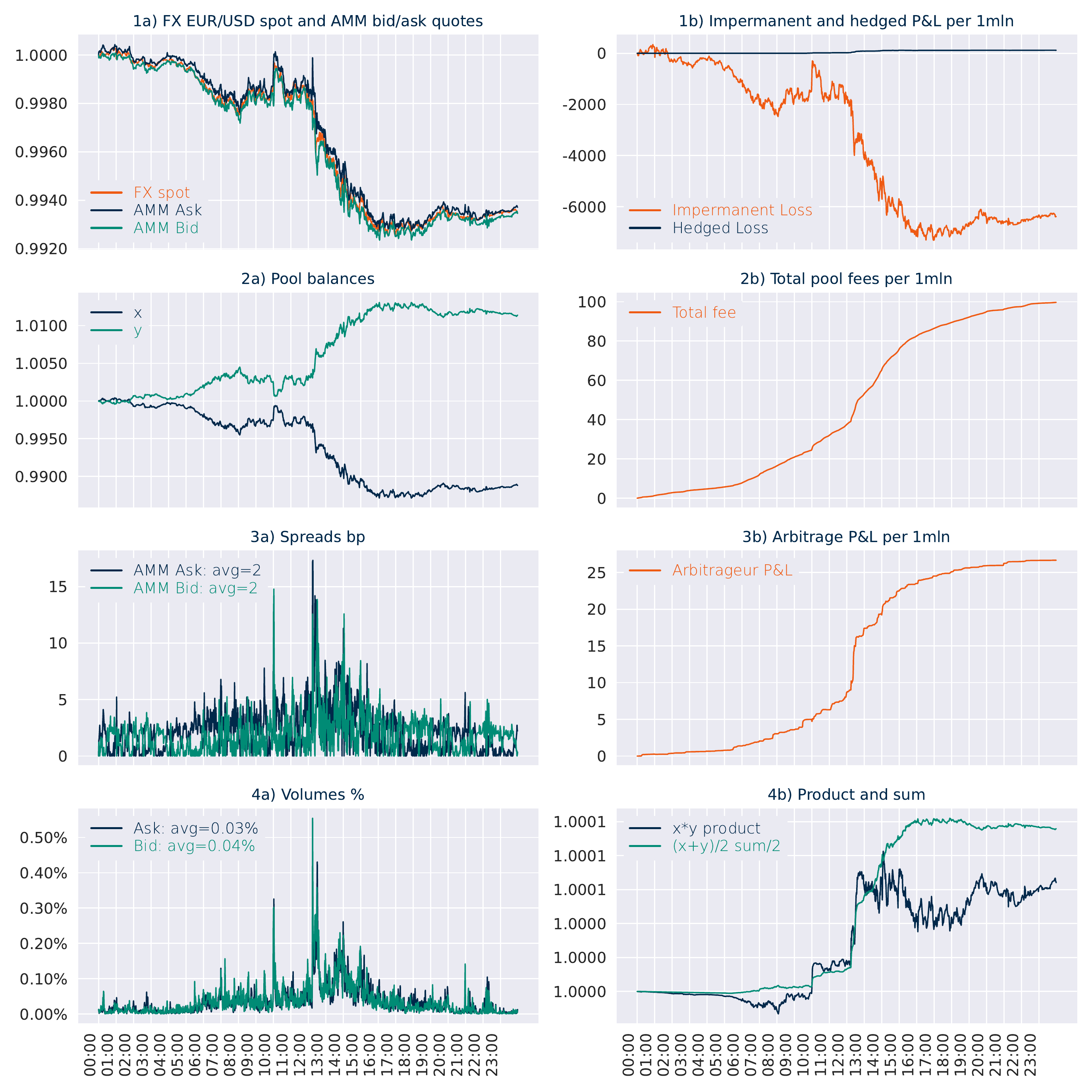}%
\vspace*{-\baselineskip}
\end{center}
\par
\vspace*{-\baselineskip}
\caption{Simulated intraday dynamics of the pool using 'EURUSD' data on 3rd
June 2021.}
\label{fig:amm_day}
\end{figure}

In Figure (\ref{fig:amm_day}), we illustrate the simulated variables from
the intraday simulation of the pool for the 'EURUSD' FX pair observed on 3rd
June 2021 using mixed product CFMM with $\alpha=5$ and fee $\epsilon=1bp$.
We show the following variables.

\begin{itemize}
\item \textbf{1a)}. The FX spot is mid-price $p^{mid}_{n}$ computed using Eq
(\ref{eq:d1}) from the market data. AMM Bid and AMM Ask are given by $%
p^{AMM, bid}_{n}$ and $p^{AMM, ask}_{n}$, respectively, computed using Eq.(%
\ref{eq:ps2a}).

\item \textbf{2a)}. Pool balances for USDC, $x_{n}$, and and EUDC, $y_{n}$,
at the end of $n$-th period computed using Eq. (\ref{eq:ps5}).

\item \textbf{3a)}. AMM Bid, $s^{AMM, bid}_{n}$, and AMM Ask, $s^{AMM,
ask}_{n}$, spreads are computed using Eq. (\ref{eq:ps2b}).

\item \textbf{4a)}. Bid, $v^{bid}_{n}$, and ask, $v^{ask}_{n}$, volumes are
computed using the given market data with Eq. (\ref{eq:d2}).

\item \textbf{1b)}. The intraday dynamics of the impermanent loss and hedged
impermanent loss of the liquidity provider of the pool corresponds to Eq. (%
\ref{eq:ps11}) and (\ref{eq:ps12}), respectively, without accounting for
total pool fees $f_{n}$.

\item \textbf{2b)}. The intraday dynamics of the total pool fees, $f_{n}$,
are computed using Eq (\ref{eq:ps10}).

\item \textbf{3b)}. The intraday dynamics of the arbitrage P\&L, $%
p\&l^{arb}_{N}$, is computed using Eq (\ref{eq:ps9}).

\item \textbf{4b)}. The product of the pool balances is computed using $%
\Pi_{n}=x_{n}y_{n}$ and the sum is computed using $%
\Sigma_{n}=(x_{n}+y_{n})/2 $ .
\end{itemize}

\section{ Analysis}

For each trading session we compute the four key realizations:

\begin{enumerate}
\item \textbf{AMM Spread in bp} is computed using Eq. (\ref{eq:ps8}).

\item \textbf{Arbitrage P\&L Annual \%} is computed using Eq. (\ref{eq:ps9})
with simple annualization of the daily p\&l equal to $260p\&l^{arb}_{N}$.

\item \textbf{Pool P\&L Annual \%} is computed using Eq. (\ref{eq:ps11})
with simple annualization of the daily p\&l equal to $260p\&l^{lp}_{N}$.

\item \textbf{Pool Hedged P\&L Annual \%} is computed using Eq. (\ref%
{eq:ps12}) with simple annualization of the daily p\&l equal to $%
260p\&l^{lp-hedged}_{N}$.
\end{enumerate}

The simulation of the pool provides path realizations of independent key
variables with the sample size of $780$, which we apply for the sensitivity
analysis. In particular, we analyze the sensitivity to the pool
specification including parameter $\alpha$ of the CFMM, pool fees $\epsilon$%
, and pool liquidity.

\subsection{ USDC/EUDC pool}

\subsubsection{ Impact of the CFMM parameter $\protect\alpha$}

\begin{figure}[]
\begin{center}
\includegraphics[width=1.\textwidth, angle=0]
{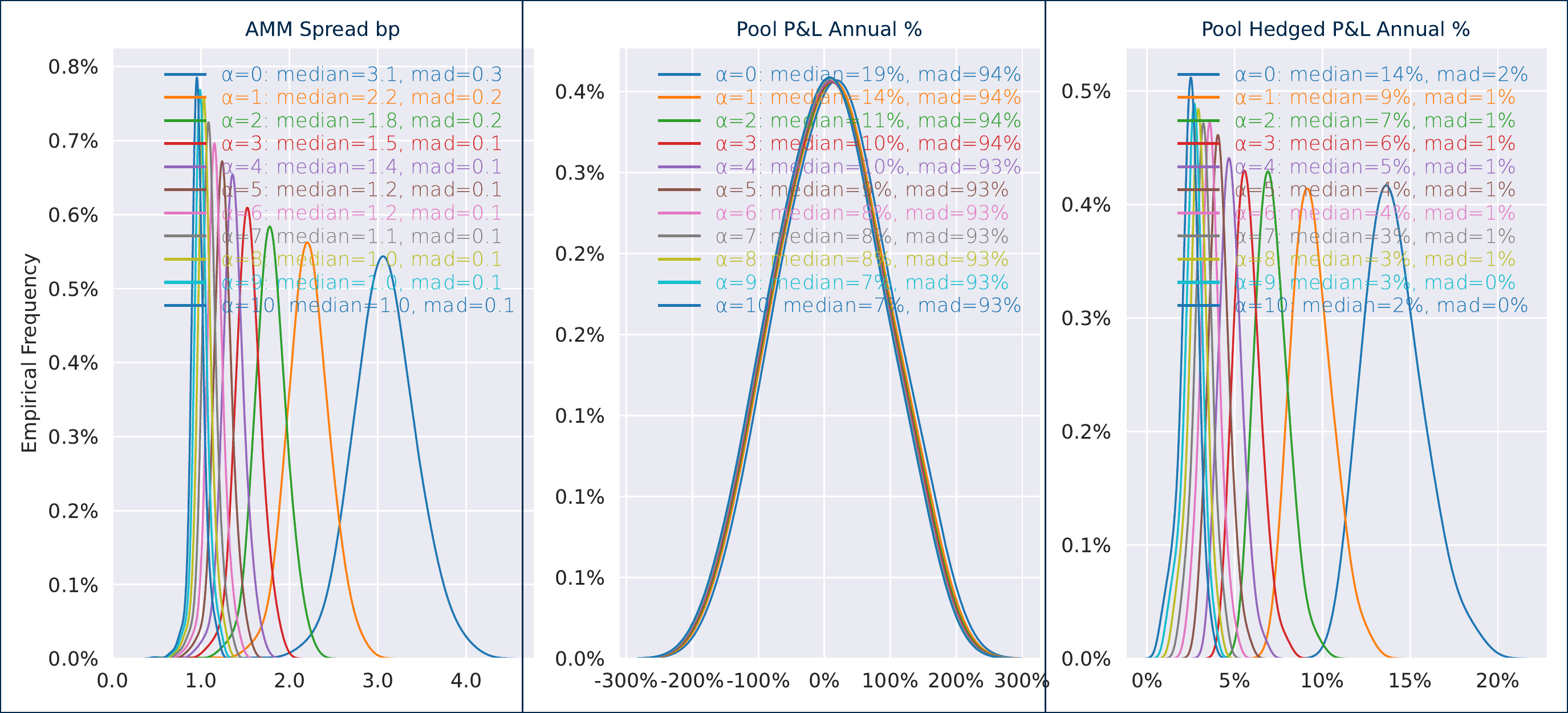}%
\vspace*{-\baselineskip}
\end{center}
\par
\vspace*{-\baselineskip}
\caption{PDFs of core pool variables for simulated USDC/EUDC pool as
functions of CFMM $\protect\alpha$, fee $\protect\epsilon=1$bp. Median is
the path median, and mad is the median absolute deviation.}
\label{fig:eur_alpha_pdf}
\end{figure}

In Figure (\ref{fig:eur_alpha_pdf}), we show the realizations of the 4 key
variables for $780$ daily realizations of the USDC/EUDC pool as functions of
the CFMM parameter $\alpha$. We show the sample statistics of the median and
median absolute deviation (mad). For $\alpha=0$, the AMM realized spreads
are the highest, while the median of the arbitrage P\&L, the pool P\&L, and
the pool hedged P\&L are the highest, and so is the deviation. As expected,
the volatility of the pool P\&L weakly depends on $\alpha$, as the price
dynamics mostly influence the P\&L. The volatility of hedged P\&L is reduced
for different levels of $\alpha$. 

\begin{figure}[]
\begin{center}
\includegraphics[width=1.\textwidth, angle=0]
{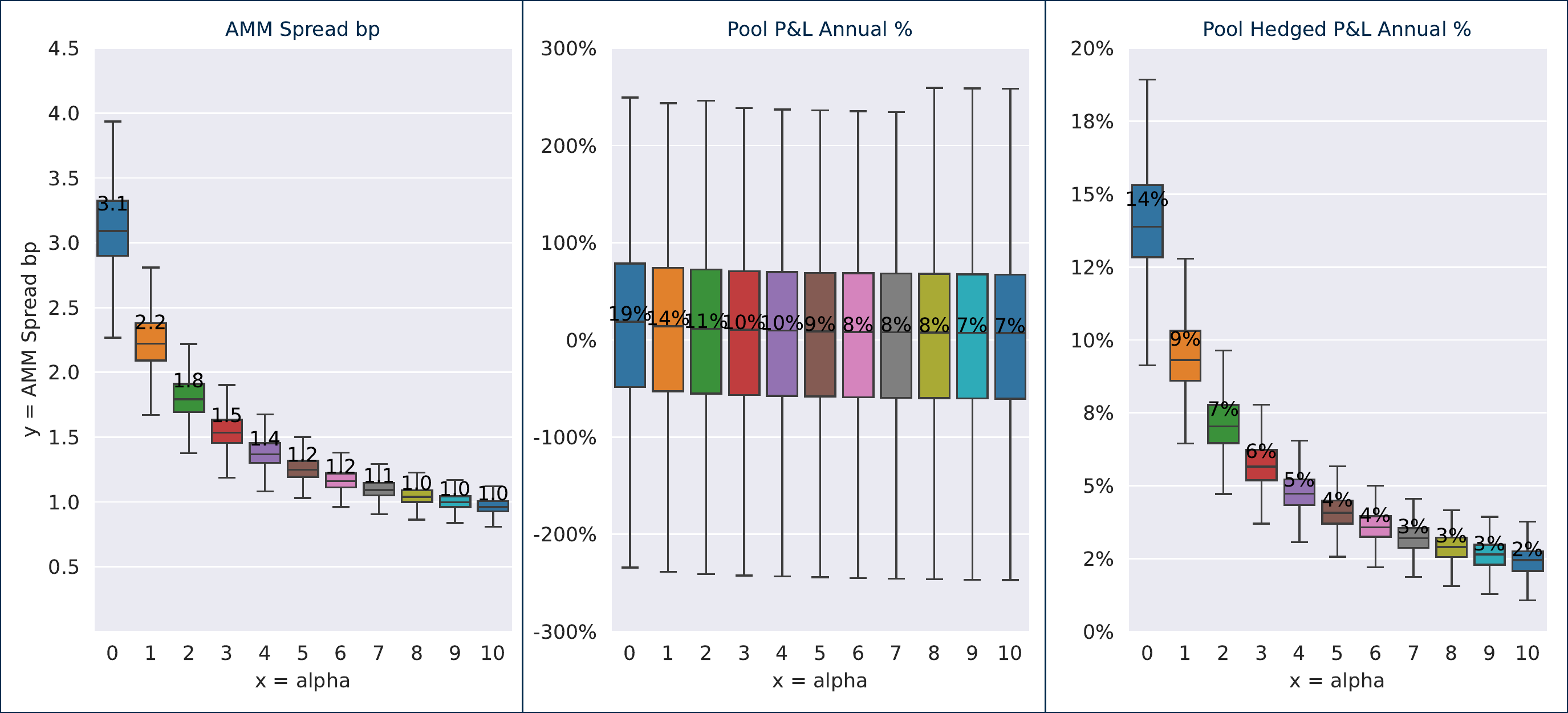}%
\vspace*{-\baselineskip}
\end{center}
\par
\vspace*{-\baselineskip}
\caption{Boxplot of core pool variables for simulated USDC/EUDC pool as
functions of CFMM $\protect\alpha$, fee $\protect\epsilon=1$bp. Median is
the path median and mad is the median absolute deviation.}
\label{fig:eur_alpha}
\end{figure}

In Figure (\ref{fig:eur_alpha}), we show the results of the same simulation
presented as the boxplot of realizations for crucial variables. From the
simulation of the AMM spread, we can select an optimal alpha that minimizes
the AMM spreads as costs to costumers and, simultaneously, maximizes the
expected P\&L for pool liquidity providers. We find an acceptable value of $%
\alpha$ equal to $5$ as an optimal trade-off that produces median AMM spread
of $3$bp, which breaks down to $1$bp of direct fees and $2$bp of the pool
liquidity costs.

\subsubsection{ Impact of end-of-day return and realized volume variance}

\begin{figure}[]
\begin{center}
\includegraphics[width=0.95\textwidth, angle=0]
{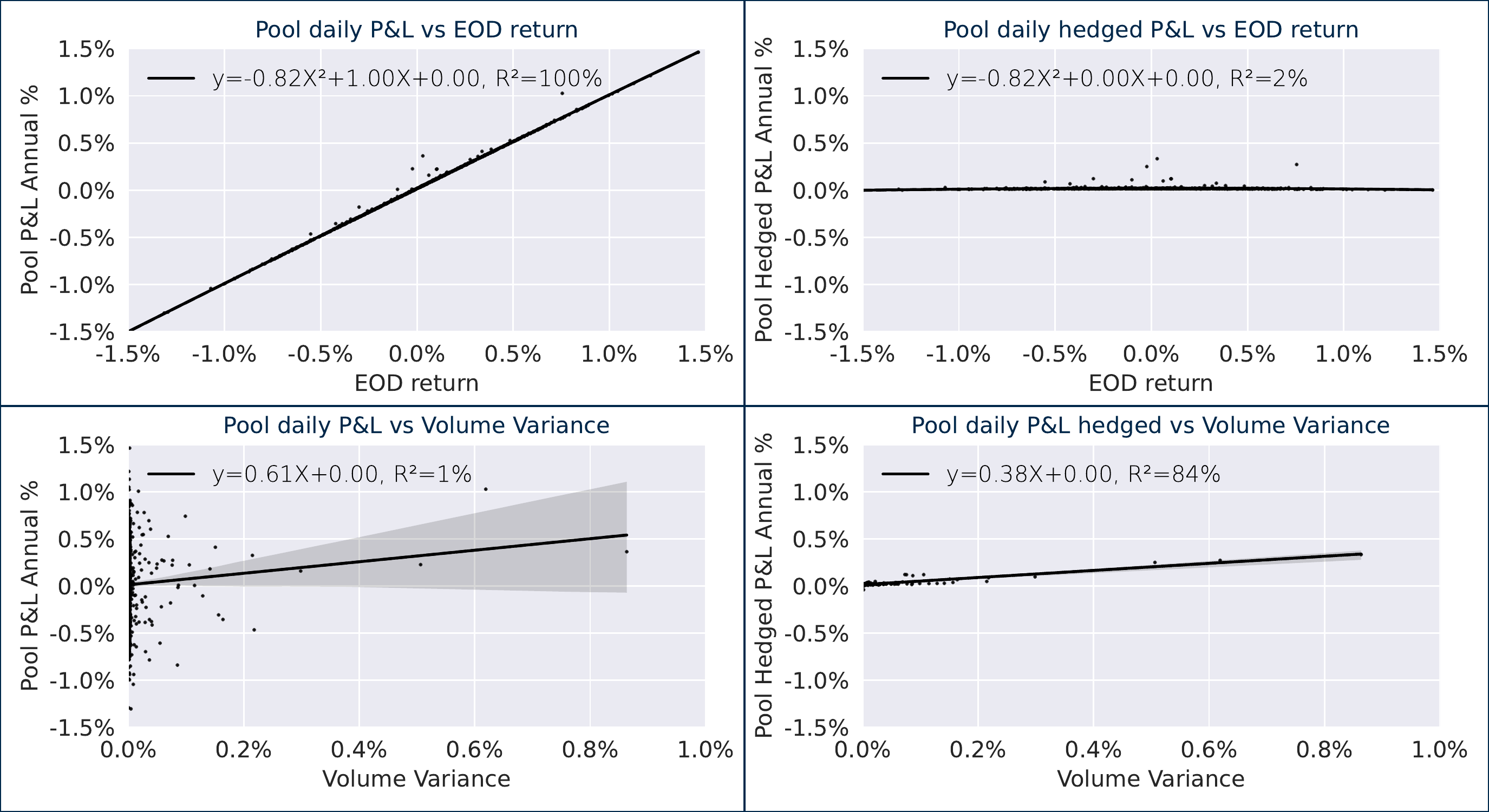}%
\vspace*{-\baselineskip}
\end{center}
\caption{The boxplot of core variables of the simulated USDC/EUDC pool as
functions of pool liquidity multiple using $\protect\alpha=5$ and fee $%
\protect\epsilon=1$bp. Median value is shown inside the bars}
\label{fig:eur_reg}
\end{figure}

In Figure (\ref{fig:eur_reg}), we show the sensitivity of daily pool P\&L
and hedged P\&L to the daily return of EURUSD mid-price and the intraday
variance of bid-ask volume. It is evident that the impermanent loss and the
daily pool P\&L has a significant sensitivity with the beta to the realized
return, with the estimate of the beta slightly above one as expected. On the
other hand, the delta-hedged pool has an insignificant beta to the realized
returns. What is also evident is that the daily P\&L, in particular, for the
hedged pool is significantly influenced by the intraday variance of bid and
ask flows.

\subsection{ G-10 pools}

\begin{figure}[]
\begin{center}
\includegraphics[width=1.\textwidth, angle=0]
{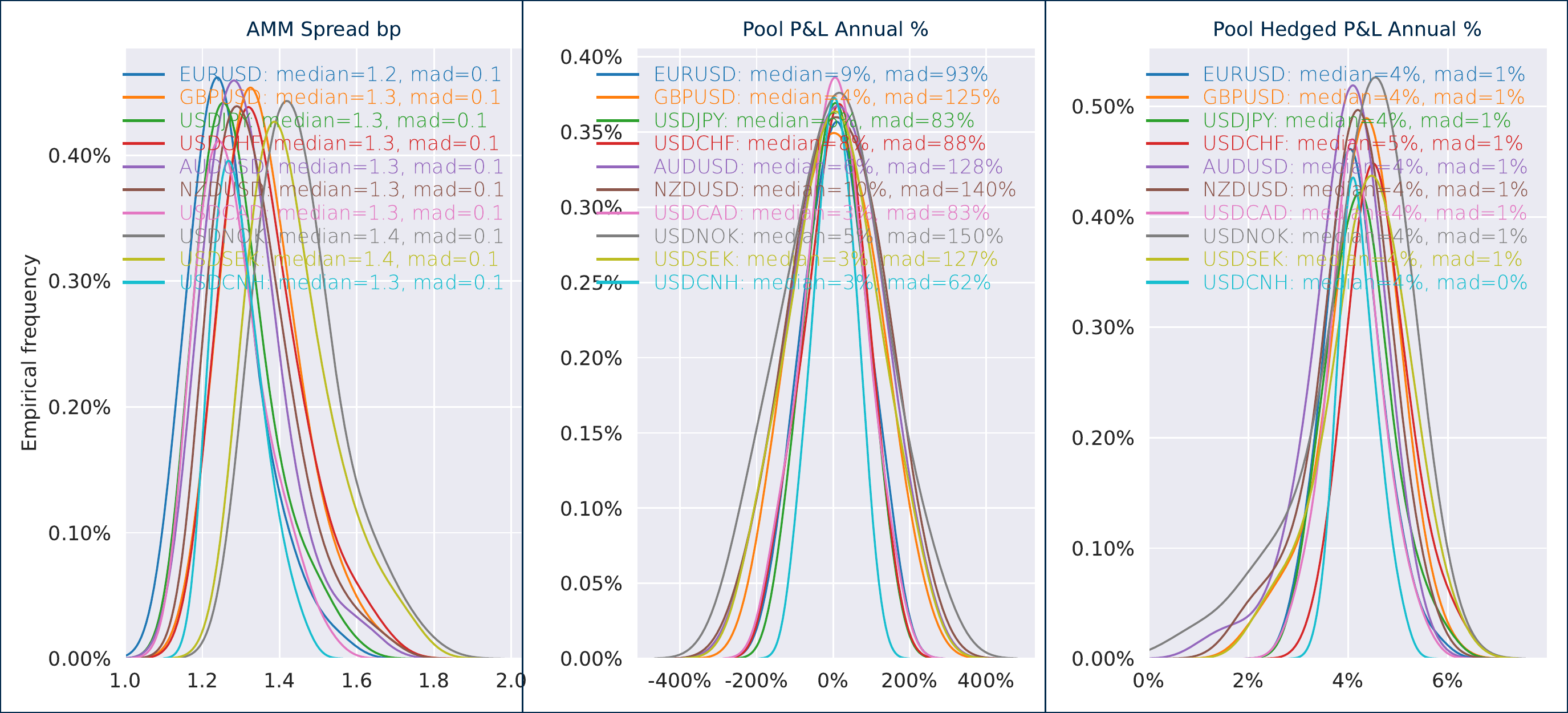} 
\includegraphics[width=1.\textwidth, angle=0]
{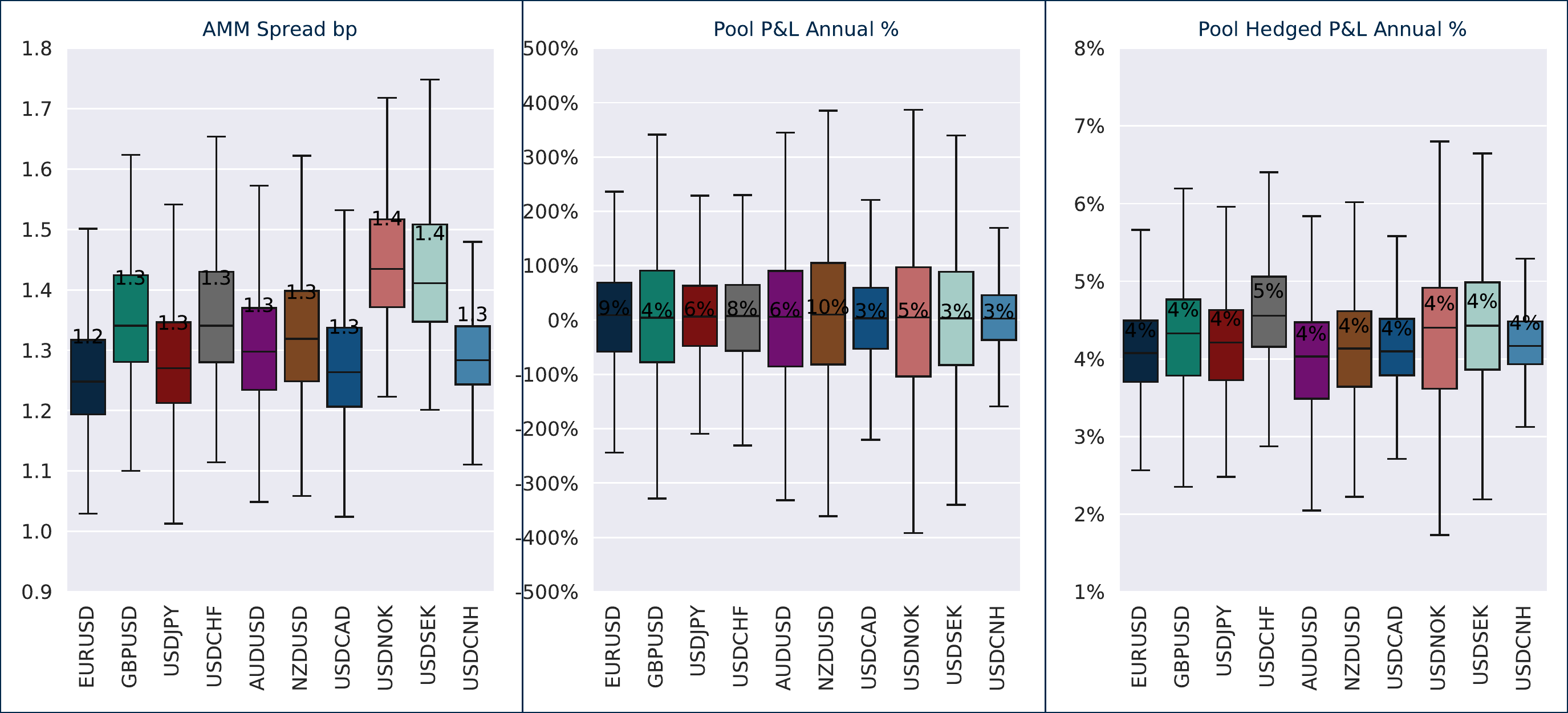}
\end{center}
\par
\vspace*{-\baselineskip}
\caption{Simulated outputs of core pool variables for G-10 universe using $%
\protect\alpha=5.0$ and fee $\protect\epsilon=1$bp. Top panel: PDFs, bottom
panel: boxplots with the median value shown.}
\label{fig:g10}
\end{figure}

In Figure (\ref{fig:g10}), we show PDFs and boxplots of the core pool
variables for the G-10 FX universe. We use $\alpha=5.0$ for the mixed
product AMM and a fee of $\epsilon=1$bp. We see that PDFs and box plots of
realized variables are close to each other. As a result, our approach is
general and can be applied to different DC pools.

\section{Conclusions}

Foreign exchange trading has been practiced since antiquity and continues to
be extremely important in our days. Canonical gospels of the New Testament
eloquently describe how Jesus expelled the money changers from the Temple.
Modern-day descendants of the money changers are known as forex market
makers. This paper described an original, novel, and powerful approach to
cleansing the Temple of high finance. Specifically, we showed that using
CaaS providers, such as Ethereum, Cardano, and Solana protocols, one can
exchange fiat currencies via AMMs, thus eliminating the need for
intermediaries.

Space does not allow us to consider exchanges of fiat currencies into
baskets of other fiat currencies, which are helpful for trading blocks
pursuing multinational endeavors such as digital trade coins. However, we
shall discuss such exchanges in a forthcoming paper.


\begin{thebibliography}{99}
\bibitem{Angeris2019} Angeris, G., Kao, H.T., Chiang, R., Noyes, C., Chitra,
T., 2019. An analysis of Uniswap markets. Cryptoeconomic Systems Journal.

\bibitem{Black1973} Black, F., and M. Scholes, 1973, The pricing of options
and corporate liabilities \emph{Journal of Political Economy} vol. 81, pp.
637-659.

\bibitem{Egorov2019} Egorov, M., 2019. StableSwap - efficient mechanism for
Stablecoin liquidity. White paper.

\bibitem{Hardjono2021} Hardjono, T., Lipton, A., and A. Pentland, 2021,
Towards an Interoperability Architecture Blockchain Autonomous Systems, 
\emph{IEEE Transactions on Engineering Management}, vol. 67, no. 4, pp.
1298-1309. Available: doi:10.1109/TEM.2019.2920154.

\bibitem{Lipton2022} Lipton, A., and T. Hardjono, 2022, Blockchain Intra-
and Interoperability, in Babich V, Birge J, Hilary G (eds), 2022, Innovative
Technology at the interface of Finance and Operations.Springer, New York.

\bibitem{Liptonetal2021} Lipton, A., A. Sardon, F. Sch\"{a}r, and C. Sch\"{u}%
pbach, 2021, Stablecoins, in Pentland A., Lipton A., and Hardjono T., 2021, 
\textit{Building the New Economy}, MIT\ Press, Boston.

\bibitem{Lipton2021} Lipton A., and A. Treccani, 2021, Blockchain and
Distributed Ledgers: Mathematics, Technology, and Economics, \emph{World
Scientific,} Singapore.

\bibitem{Nakamoto2008} Nakamoto, S., 2008. Bitcoin: A peer-to-peer
electronic cash system. Available: https://bitcoin.org/bitcoin.pdf.

\bibitem{Schar2021} Sch\"{a}r, F., 2021. Decentralized finance: On
blockchain-and smart contract-based financial markets. FRB of St. Louis
Review.

\bibitem{WBG2021} World Bank Group, 2021, Blockchain Interoperability.
Available: %
\url{http://documents.worldbank.org/curated/en/373781615365676101/Blockchain-Interoperability}
\end{thebibliography}
\end{document}